\documentclass[twocolumn]{emulateapj}
\usepackage{graphics, enumerate, graphicx, color}
\usepackage{psfig, epsfig}
\usepackage{natbib}

\newcommand{\bearr}{\begin{eqnarray}}
\newcommand{\eearr}{\end{eqnarray}}
\definecolor{titlecol}{rgb}{0,0,0}

\def\changed    {\color{titlecol} }

\def\csqnu              {$\chi^2_{\nu}$}

\newcommand{\eqnstart}   {\begin{equation}}
\newcommand{\eqnend}     {\end{equation}}

\def\BT		{bulge-to-total}

\def\MBHminr    {2^{+19}_{-1.5} \times 10^5 M_{\odot}}

\def\MBHmaxr    {2^{+21}_{-1.8} \times 10^9 M_{\odot}}

\def\MBHavgr    {3 \times 10^7 M_{\odot}}

\def\MBHmedr    {4 \times 10^7 M_{\odot}}

\def\galfit     {{\tt GALFIT}}
\def\mgii		{$\mathrm{Mg \textsc{ii}}$}

\newcommand{\msol}{\rm M_\odot}

\slugcomment{Submitted to Ap. J.}

\shorttitle{Moderate-Luminosity AGN at $z \sim 2$: Varied Accretion in Disk-Dominated Hosts}
\shortauthors{Simmons et al.}

\begin{document}

\title{Moderate-luminosity Growing Black Holes from $1.25 < z < 2.7$: Varied Accretion in Disk-Dominated Hosts}

\author{B. D. Simmons$^{1, 2, 3}$, C. M. Urry$^{2, 4}$, K. Schawinski$^{2, 4, 5}$, C. Cardamone$^{6}$, E. Glikman$^{1,2,}$\altaffilmark{7}} 
\affiliation{1 Astronomy Department, Yale University, New Haven, CT 06511, USA}
\affiliation{2 Yale Center For Astronomy \& Astrophysics, Physics Department, Yale University, New Haven, CT 06511, USA}
\affiliation{3 Department of Physics, University of Oxford, Keble Road, Oxford OX1 3RH, UK}
\affiliation{4 Physics Department, Yale University, New Haven, CT 06511, USA}
\affiliation{5 Department of Physics, Institute for Astronomy, ETH Zurich, Wolfgang-Pauli-Strasse 16, CH-8093 Zurich, Switzerland}
\affiliation{6 Brown University, Box 1912, 96 Waterman St., Providence RI  02912, USA}
\altaffiltext{7}{NSF Fellow}

\email{brooke.simmons@astro.ox.ac.uk}

\begin{abstract}
We compute black hole masses and bolometric luminosities for 57 active galactic nuclei (AGN) in the redshift range $1.25 \leq z \leq 2.67$, selected from the GOODS-South deep multi-wavelength survey field via their X-ray emission. We determine host galaxy morphological parameters by separating the galaxies from their central point sources in deep \emph{HST} images, and host stellar masses and colors by multi-wavelength SED fitting. 90\% of GOODS AGN at these redshifts have detected rest-frame optical nuclear point sources; bolometric luminosities range from $2 \times 10^{43-46} \mathrm{~erg~s}^{-1}$. The black holes are growing at a range of accretion rates, with {\changed $\gtrsim 50$\%} of the sample having $L/L_{Edd} < 0.1$. 70\% of host galaxies have stellar masses $M_* > 10^{10} M_{\odot}$, with a range of colors suggesting a complex star formation history. We find no evolution of AGN bolometric luminosity within the sample, and no correlation between AGN bolometric luminosity and host stellar mass, color or morphology. Fully half the sample of host galaxies is disk-dominated, with another 25\% having strong disk components. {\changed Fewer than} $15$\% of the systems appear to be at some stage of a major merger. These moderate-luminosity AGN hosts are therefore inconsistent with a dynamical history dominated by mergers strong enough to destroy disks, indicating minor mergers or secular processes dominate the co-evolution of galaxies and their central black holes at $z \sim 2$.
\end{abstract}


\pagebreak

\section{Introduction}

The black hole-galaxy connection is well-established at local \citep[e.g.,][]{ferrarese99,gebhardt00,marconi03,haringrix04} and moderate \citep[e.g.,][]{treu04,woo05,peng06,bennert10,bennert11a} redshifts. {\changed Clues to understanding the co-evolution of galaxies and their central supermassive black holes come from detailed study of both moderate-luminosity growing black holes and the galaxies that host them. It is these moderate-luminosity AGN and host galaxies that cumulatively dominate the observed X-ray background and thus constitute a large fraction of the underlying black hole growth in the universe \citep{hasinger05,treisterurry11}.}

Examination of both AGN and host galaxies requires separation of their blended signal, preferentially with a combination of space- and ground-based data at multiple wavelengths. Using deep \emph{Hubble Space Telescope} (\emph{HST}) data to extract nuclear point sources from AGN+host images requires care but provides the leverage needed to determine AGN bolometric (total) luminosities and host galaxy stellar masses and morphologies \citep{simmons08, simmons11}.

Until recently, such analyses for large, uniform AGN+host samples in the rest-frame optical were only possible to $z \lesssim 1.3$ using deep survey data taken with the Advanced Camera for Surveys (ACS) on \emph{HST}. Results of multiple studies of host galaxy morphologies and black hole growth rates at those redshifts indicate a complex picture, with moderate-luminosity AGN predominantly powered by high-mass, slow-growing black holes \citep{simmons11} in host galaxies with a range of morphologies \citep{sanchez04,bundy08,gabor09}. 

Mergers are often invoked to trigger black hole accretion \citep[e.g.,][]{sanders88,croton06,hopkins06}, but mergers are equally common in galaxies hosting actively growing and inactive black holes \citep{grogin05}. Moreover, moderate-luminosity AGN at $z\sim 1$, which are powered by high-mass black holes that have already undergone significant growth, commonly found in disk-dominated systems \citep{simmons11}. This raises the issue of whether black hole growth is similar at $z \sim 2$, where star formation \citep{lilly96,madau98} and AGN activity \citep[e.g.,][]{fan01,wall05} both peak, or whether mergers are relatively more important.

The recently-installed Wide Field Camera-3 (WFC3) instrument on \emph{HST} now provides high-resolution, rest-frame optical images of AGN and host galaxies to $z \sim 2$. With the Cosmic Assembly Near-infrared Deep Extragalactic Legacy Survey \citep[CANDELS;][]{grogin11} imaging in the deep \emph{Chandra} legacy fields, we can separate AGN and hosts at rest-frame optical and UV wavelengths, enabling detailed investigations of the properties of both moderate-luminosity growing black holes and their host galaxies.

In Section \ref{data} we describe our sample selection and analysis methods. Section \ref{results} presents the AGN host galaxy morphologies, stellar masses and colors, as well as estimates of black hole masses and growth rates for the AGN in the sample. Section \ref{discussion} discusses these results in the overall context of black hole-galaxy co-evolution, specifically merger-driven versus secular growth scenarios.

Throughout this paper, we adopt a $\Lambda$CDM cosmology with $H_0 = 71 \mathrm{~km~s}^{-1}\mathrm{~Mpc}^{-1}, \Omega_M = 0.27, \Lambda_0 = 0.73$ \citep{spergel03}.

\section{Data}\label{data}

\subsection{Sample Selection}

The \emph{Chandra} Deep Field - South is one of the best-observed parts of the sky, with deep multi-wavelength coverage from both ground- and space-based observatories. In particular, the 4 Megaseconds of combined X-ray observations within the field \citep{xue11} allow for selection of AGN at $z \sim 2$ that is relatively unbiased with respect to obscuration. Examination of those AGN and their hosts at rest-frame optical wavelengths is possible with the multi-epoch deep observations of the GOODS-South field \citep[chosen to coincide with the CDF-S; see][]{giavalisco04} within the CANDELS survey \citep{grogin11}. We match the X-ray sources to WFC3 sources using a maximum-likelihood method \citep{cappelluti09,cardamone10} that considers both positional overlap and uncertainties as well as multi-wavelength source fluxes to determine the most probable optical match to each X-ray source. 

We select AGN within the \emph{Chandra}-GOODS-S-CANDELS field according to the following criteria: 
\begin{enumerate}
\item Redshifts \citep[compiled by][]{xue11} within the range $1.25 < z < 2.67$, so that the observed wavelength ranges spanned by the F125W and F160W WFC3 \emph{HST} filters are in the rest-frame \emph{B} band;

\item Source type identification as AGN within the \citet{xue11} 4 Ms X-ray catalog. Briefly, AGN are identified via X-ray luminosity, photon index, X-ray-to-optical flux ratio, X-ray excess over measured star formation, and/or optical spectroscopic features. The criteria are described further in \citeauthor{xue11}; 

\item Extended source detection within the WFC3 image with a signal-to-noise of at least 5 per pixel to ensure robustness of morphological fitting; and

\item Extended source detections in at least five optical and infrared bands {\changed \citep[including WFC3 \emph{JH} from CANDELS as well as ACS \emph{BVIz} and ground-based \emph{Ks} from GOODS; details of GOODS observations are given in Tables 1 and 2 in][]{giavalisco04}}, to facilitate accurate determination of stellar masses and rest-frame colors.
\end{enumerate} 

Of the 57 X-ray-selected AGN in the sample using the above criteria, {\changed 26 of the redshifts are spectroscopic \citep{szokoly04,zheng04,mignoli05,vanzella08,popesso09,silverman10} and 31 are photometric \citep{luo10,cardamone10,rafferty11}.} 41 are considered luminous AGN by \citeauthor{xue11} owing to their having absorption-corrected, rest-frame $0.5-8$~keV luminosities $L_X \geq 3 \times 10^{42}$~erg~s$^{-1}$. An additional 14 sources have $10^{42} \leq L_X < 3 \times 10^{42}$~erg~s$^{-1}$, and the remaining two have luminosities of $9.1$ and $9.5 \times 10^{41}$~erg~s$^{-1}$. The median and mean X-ray luminosities of the sample are $4.4$ and $7.3 \times 10^{42}$~erg~s$^{-1}$, respectively, and the most luminous X-ray AGN in the sample has $L_X = 2.6 \times 10^{44}$~erg~s$^{-1}$, consistent with expected detections within the volume of GOODS-S based on the AGN luminosity function \citep{croom04}.

The CANDELS survey \citep{grogin11} includes several deep and wide fields on the sky with both the WFC3 and ACS cameras on \emph{HST}. We perform morphological fits on the publicly available co-added six-epoch images in both the F125W and F160W bands, which were drizzled to a resolution of $0''.06$ per pixel. For details of the image processing, see \citet{koekemoer11}. 

Morphological decomposition of two-dimensional AGN+host galaxy images requires accurate determination of the point-spread function (PSF). We determine the PSF of the F125W and F160W CANDELS images using the PSF-modeling routines in the IRAF package {\tt daophot}. This method balances the noise-free attributes of a simulated PSF model with the detailed particulars of the actual PSF of multi-drizzled, multi-epoch data. We verified that the WFC3 PSF is very stable across the deep mosaic in the GOODS-S field, so we determine a single PSF for each band and use it in the morphological analysis of each of the sources.

\subsection{Morphological Fitting}

We perform two-dimensional parametric morphological fitting on the F125W and F160W images for the 57 sources using \galfit\ \citep{peng02}, which is very reliable for recovering both normal galaxy morphologies and AGN-host decomposition \citep[see][for relevant simulations]{haussler07,simmons08,pierce10a}.

Our detailed fitting procedure is similar to that performed in previous simulations and studies of AGN host morphologies \citep{simmons08, simmons11}. We determine initial guesses for fit parameters such as total source magnitude, half-light radius, axis ratio, position angle and centroid position using SExtractor \citep{bertin96}. The value of the sky background varies slightly across the CANDELS mosaic, so we fix the SExtractor-determined value for each source and each iteration of the morphological fitting in order to avoid the confusion of sky background and extended source light that can ensue when the sky background is allowed to vary along with the other fit components.

We fit each galaxy with two models in each band: (1) a host galaxy modeled by a S\'ersic profile \citep{sersic68}, and (2) a host galaxy with a central point source. We determine the final fit parameters via an iterative process that initially fits only the central portion of the AGN+host, to find the best-fit centroid positions of each fit component. Those positions are then held fixed in successive iterations, which zoom outward to eventually encompass a large enough image region such that the extended portions of the central source are well sampled. Where necessary, we fit nearby, bright and/or extended companions in order to simultaneously model their extended light distributions. Companion fitting (and masking of fainter, more compact companions further from the central source) ensures a more realistic fit to the central source. The first three fits to each source are performed with the use of a customized adaptive batch-fitting code. Each source is then followed up individually to ensure the true minimum reduced \csqnu\ value within the parameter space is achieved.

For each source and each fit (S\'ersic only and S\'ersic+Point Source), we perform one final iteration where every parameter (with the exception of the sky background) is allowed to vary. This final iteration ensures that \galfit\  has found the true best fit and also allows for an estimate of the computational uncertainties for each parameter, including the centroid positions of each component. For all sources, the differences between the final fit parameters and those of the penultimate iteration are well within the reported parameter uncertainties; thus, we are confident that we have found the true best-fit parameters for each source and fit type. Typically, an AGN+host galaxy in the sample requires 6-8 total iterations for a S\'ersic+PSF fit, with 1-4 fitted companions. Some of the sources require additional iterations, usually due to the presence of multiple (occasionally more than 4) bright, nearby companions.

\begin{figure}
\figurenum{1}
\plotone{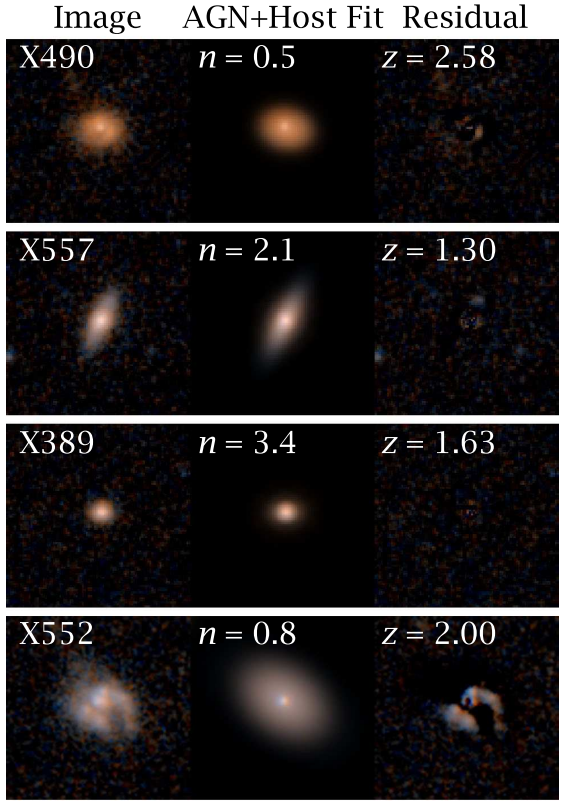}\label{fitsample}
\caption{
\emph{HST} WFC3 $F125W+F160W$ images (left), 2-dimensional morphological fits (center), and residuals (image - fit; right) for four X-ray selected AGN+hosts in this sample (from top to bottom): a disk-dominated, intermediate, and strong-bulge host morphology, as well as a clumpy/asymmetric host galaxy. Object 389 does not have a detected nuclear point source; the other three have point sources {\changed that are faint but detected at} $\geq 5 \sigma$.
}
\end{figure}

The final step in the determination of AGN+host morphologies involves comparing the S\'ersic-only and the S\'ersic+Point Source fits for each of the 57 sources in order to determine whether we significantly detect a central point source. Detailed simulations of the recovery of central point sources within ACS data \citep{simmons08} suggest that failed detections of intrinsically-detectable point sources (\emph{i.e.}, false negatives) are rare, but spurious point-source detection (\emph{i.e.}, false positive) rates are strongly dependent on host morphology, ranging from $\sim 1\%$ in a sample of pure disk galaxies to $\sim 25\%$ for a sample of pure bulges. When \galfit\ does not detect a central point source, therefore, this is most likely a robust indicator that any light from a central AGN is below the detection threshold of the image. Detection of a central point source in a S\'ersic+Point Source fit, on the other hand, is not a guarantee of its authenticity.

We therefore assess the robustness of each point-source detection based on a number of criteria. First, if an F-test indicates that the addition of a point source significantly improves the \csqnu\ goodness-of-fit parameter, we consider the point source genuine. Alternatively, we consider a point source genuine if the S\'ersic-only fit converges to an unrealistically high value for the S\'ersic index $n$ {\changed\citep*[generally galaxies have $n \lesssim 10$; e.g.,][]{caon93}}. Lastly, we examine the central portion of the fit residuals. We consider a point source genuine if the residual of the S\'ersic-only fit shows obvious signs of additional light from an unresolved source (however, in practice at least one of the other two criteria also apply if this last criterion is true).

Figure \ref{fitsample} shows a selection of fitted AGN+hosts; we present the full catalog of best-fit morphological parameters for each source in Table~\ref{datatable}. The uncertainties reported in Table~\ref{datatable} are determined by combining in quadrature the reported computational uncertainties from \galfit\ and the uncertainties in morphological parameter recovery determined from the simulation of over 50,000 AGN host galaxies by \citet{simmons08}. The latter typically dominate the overall uncertainty in a morphological fit.

\section{Results}\label{results}

\subsection{AGN Host Morphologies}\label{sec:morph}

Just under half (25) of the host galaxies in this sample are best described as single-component galaxies with smooth residuals after subtracting a single-component host fit. Fourteen have clumpy morphologies, with multiple offset components either embedded in a larger component or very closely separated. Eighteen host galaxies ($\approx 32\%$) show signs of asymmetric features or tidal tails in the images and/or fit residuals; 8 of these appear to be involved in major mergers, consistent with an independent assessment of an overlapping sample using visual morphologies \citep{kocevski11}. 

Using the F-test to assess the significance of improvements to the fit with the addition of a central point source, we find that $90\%$ of AGN+hosts have point source detections with at least $3 \sigma$ significance. The majority are at least $5 \sigma$ point-source detections. We quote point-source-subtracted host morphologies for the 51 sources that have a point source at a minimum significance level of $3 \sigma$; for the remaining six sources we quote the morphology from the S\'ersic-only fit. As expected, for those sources where the addition of a point source was not significant, the host morphology does not significantly change between the S\'ersic only and the S\'ersic + point source fit. 

\begin{figure}
\figurenum{2}
\epsscale{1.15}
\plotone{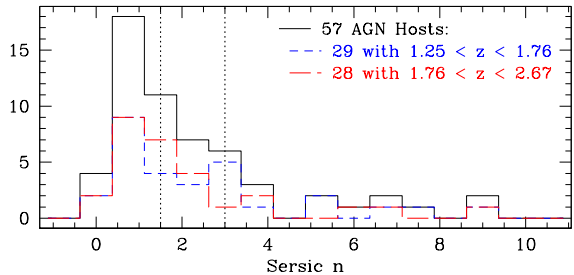}\label{nhist}
\caption{
Distribution of fitted S\'ersic indices for the sample of 57 AGN host galaxies at $1.25 < z < 2.67$. The sample is characterized by disky morphologies (median $n = 1.5$). This is conservative: extensive simulations of AGN host morphologies \citep{simmons08} show that while a fitted $n \lesssim 1.5$ strictly confines the bulge contribution to less than $20\%$, AGN host galaxies with fitted $n = 4$ can still have significant intrinsic disks. We follow \citeauthor{simmons08} in separating host galaxies into three groups based on cuts at $n=1.5$ and $3$ (dotted lines). S\'ersic indices for sub-samples in two redshift bins spanning equal time intervals $\left(\Delta t \approx 1.3 \mathrm{~Gyr}\right)$ are consistent with being drawn from the same parent sample.
}
\end{figure}

We show the fitted S\'ersic indices for the AGN host galaxies in Figure \ref{nhist}. Morphologies of the hosts, or (for the clumpy hosts) of the component within the host associated with the nuclear source, are typically disky. Just over half the sample has $n < 1.5$ (29 of 57 hosts), an unambiguously disk-dominated morphology. An additional 15 have $1.5 \leq n < 3$, which is consistent with intermediate morphologies (bulge-to-total ratios between 20 and $65\%$). Indeed, \citet{simmons08} have shown that even galaxies fit with $n=4$ can have a significant disk component (up to $45\%$). Five sources have $3 < n < 4$, and 8 have $n > 4$. The overall sample can therefore be characterized as having a mix of morphological types, but disks are present in roughly three-quarters of the sample, and $\sim 50\%$ of hosts are consistent with having a bulge fraction of less than 20\%. (In fact, 21 of the 29 disk-dominated sources have $n \leq 1$, consistent with a \BT\ ratio of $ \leq 10\%$.)

The redshift range over which the $F125W$ and $F160W$ WFC3 filters sample the rest-frame $B$ band spans approximately 2.6~Gyr of cosmic time. Splitting the sample into two sub-samples of $\Delta t \approx 1.3\mathrm{~Gyr}$ each reveals no difference in the distribution of morphologies between the two bins, according to a Kolmogorov-Smirnov (K-S) test.

\subsection{Host Galaxy Stellar Masses and Colors}

We determine stellar masses of the host galaxies using the multi-wavelength {\changed GOODS+CANDELS $BVIzJHKs$ SEDs. We minimize the effect of the AGN on the host-galaxy SED fitting by first subtracting the point-source contribution to the flux in each \emph{HST} band before fitting the SEDs with the stellar population synthesis templates of \citet{maraston05} using FAST \citep{kriek09a}. 

Subtracting the point sources in the \emph{BVIzJH} bands generally has only a small effect on the recovered galaxy masses and colors due to the fact that the host galaxies dominate the observed SEDs: the median $L_{\rm host} = 10 \times L_{\rm AGN}$ in $JH$. However, the effect is systematic: because using removing the AGN flux lowers the source flux used to calculate mass, the recovered stellar mass is lower than it would be were the presence of the AGN not accounted for. Additionally, for a sample such as this, where there is little bias against obscured AGN due to hard X-ray selection, but unobscured AGN are selected against by the requirement that the host galaxy morphology be recoverable, most nuclear point sources are reddened. Removing the AGN flux from the SED therefore tends to recover a slightly bluer color than would be calculated if the effect of the AGN were neglected. This is likewise a small but systematic effect. 

Once the point-source SED is subtracted from the \emph{HST} bands, extrapolating it to remove an estimated contribution to the \emph{Ks}-band flux makes no difference within the uncertainties. We therefore subtract the nuclear point-source contribution to the SED in the \emph{HST} bands where we have enough angular resolution to confidently separate AGN from host galaxy, and refrain from unnecessary extrapolation of the SED to the ground-based \emph{Ks} band.}

\begin{figure}
\figurenum{3}
\epsscale{1.15}
\plotone{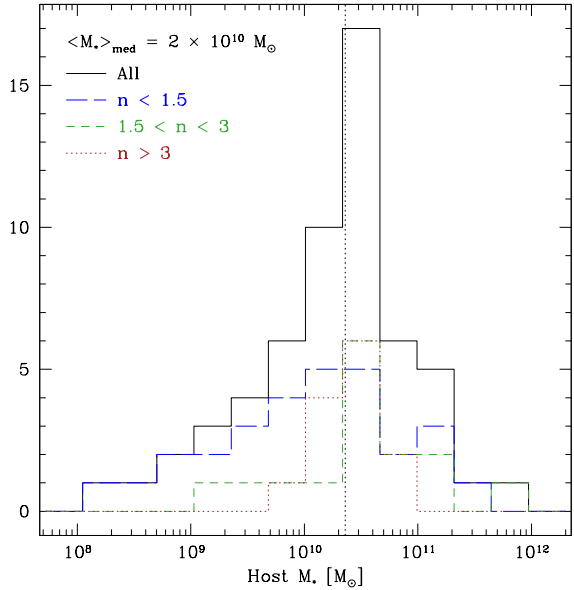}\label{Mhist}
\caption{
Histogram of stellar masses for AGN hosts with $1.25 < z <  2.67$. The full sample (solid black line) has a median value of $2  \times 10^{10}$ $M_{\odot}$ (dotted line); we also show histograms for the  unambiguously disk-dominated ($n < 1.5$; blue long-dash line),  intermediate ($1.5 < n < 3$; green short-dash line), and  strong-bulge ($n > 3$; red dotted line) sub-samples. Disk-dominated host galaxies tend to have lower masses than those with strong-bulge morphologies (K-S $\approx 93\%$).
}
\end{figure}

To test the dependence of recovered stellar masses on the choice of template, we {\changed compared stellar masses calculated using the templates of \citet{bc03} to} the masses calculated from \citet{maraston05} templates; we found that the recovered stellar mass is not strongly dependent on the choice of template set. The distribution of derived masses is shown in Figure \ref{Mhist}. The minimum and maximum masses of the sample are $2 \times 10^8$ and $6 \times 10^{11}$ $M_{\odot}$, respectively; the average and median masses are $1.7 \times 10^{10}$ and $2.3 \times 10^{10}$ $M_{\odot}$, respectively. $70\%$ of the sample has $M_* > 10^{10}$ $M_{\odot}$, which is expected since AGN are typically found in massive hosts \citep{cardamone10}; this probably reflects a selection effect missing slow-growing black holes in low-mass galaxies rather than an intrinsic distribution \citep{aird12}. The average uncertainty in the mass is $0.8$ dex.
 
\begin{figure}
\figurenum{4}
\epsscale{1.15}
\plotone{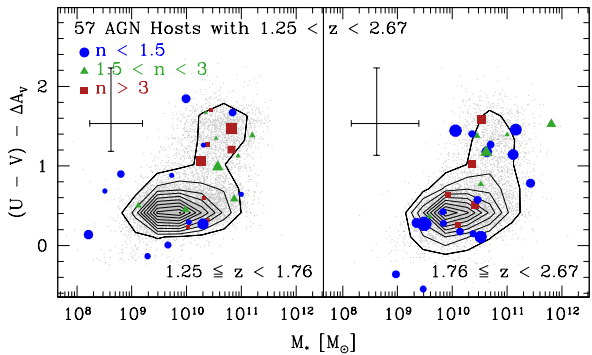}\label{cmd}
\caption{
Rest-frame $(U-V)_{AB}$ color versus stellar mass for the host galaxies of AGN with $1.25 < z < 2.67$, separated into two redshift bins of approximately equal $\Delta t \approx 1.3 \mathrm{~Gyr}$. Disk-dominated host galaxies (blue circles) populate both the ``blue cloud'' and ``red sequence'' portions of the diagram. Strong-bulge hosts (red squares) are more likely to have redder colors, but not exclusively. The stellar mass distributions are statistically indistinguishable, {\changed and the dust-corrected color distributions of AGN host galaxies between the two redshift bins are only different at the 75\% level, a marginal difference at best}. Median error bars are shown in each panel. The size of each point representing a host galaxy is proportional to its AGN bolometric luminosity. Contours and gray points show the positions of inactive galaxies in the same redshift bins \citep{whitaker11}.  
}
\end{figure}

We visually inspected each of the best-fit templates from FAST and verified the goodness-of-fit. We use the best-fit templates to determine the rest-frame $\left(U-V\right)$ colors of each host galaxy, following \citet{brammer09} in assuming a \citet{calzetti00} dust law to correct for dust, such that $\left(U-V\right)_{corr} = \left(U-V\right) - \Delta A_V$, where $\Delta A_V = 0.47 A_V$. We show the color-mass diagrams for the AGN host sample in Figure \ref{cmd}, compared to inactive galaxies with the same redshifts \citep[from the NEWFIRM medium-band survey; ][]{whitaker11}. We separate the sample into two redshift bins divided at $z = 1.76$, so that each bin represents a time interval of $\Delta t \approx 1.3 \mathrm{~Gyr}$. 

A K-S test indicates no significant difference between the stellar mass distributions of the AGN hosts in the two redshift bins. The uncorrected $U-V$ colors between the bins are inconsistent with being drawn from the same parent sample at the 97\% level, but the significance is reduced to 75\% ($\sim 1 \sigma$) once the dust correction is applied: correcting for dust shifts more galaxies into the blue cloud in the lower redshift bin than in the higher. {\changed A detailed comparison} of the colors of morphological sub-samples in different redshift bins is highly uncertain due to small numbers in each sub-sample. {\changed However, we observe intrinsically red disk-dominated hosts in each bin (at 85\% confidence given the uncertainties), as well as blue strong-bulge hosts (99\% confidence).}

Note that, although the stellar mass determination is unaffected (within the uncertainties) by the choice of model templates, the value of the dust correction $A_V$ can vary significantly depending on the choice of template. We include differences in each source's dust-correction between the \citet{bc03} and \citet{maraston05} templates in the uncertainties shown in Figure \ref{cmd}, noting that the choice of template does not change the overall result.

\subsection{Black Hole Masses and Accretion Rates}\label{bhmass}

Black hole mass, one of the fundamental properties of an AGN, can be measured directly only for very nearby sources. If a black hole is unobscured, measurements of broad spectral lines from the region near the black hole allow for calculations of central masses via the virial method. However, this is not generally possible for more obscured sources like those that dominate the present sample{\changed : only 1 AGN in this sample has a measurable broad line (see below), so this method is not feasible for the entire sample.} Instead, we use well-characterized relations between host galaxy properties and black hole masses, which allow estimates of black hole masses out to high redshift as long as evolutionary corrections are considered.

Locally, black hole mass correlates with the velocity dispersion \citep[\emph{e.g.,}][]{ferrarese00, gebhardt00}, mass \citep[\emph{e.g.,}][]{magorrian98, haringrix04} and optical luminosity \citep[\emph{e.g.,}][]{marconi03,ferrarese05,graham07} of the stellar bulge. These relations may evolve as a result of multiple effects \citep{treu04,borys05,woo05,peng06,alexander08, woo08,jahnke09,decarli10b,merloni10,bluck11}. It appears that bulge stellar populations are younger at earlier times, and it also appears that bulge growth may trail behind black hole growth at early times. However, the evolutionary corrections are highly uncertain. Recent work \citep{cisternas10, merloni10} suggests that the relationship between the central black hole mass and the \emph{total} stellar mass of the host galaxy does not evolve to $z \sim 2$. Other groups find that the relation evolves by a factor of $\sim 2-4$ to $z \sim 2$ \citep{trakhtenbrot10,bennert11b}, but the highest-redshift measurement {\changed to date} of a single black hole with spatially-resolved spectroscopy \citep{inskip11} is consistent with the local black hole-galaxy relation.

There are therefore many sources of uncertainty in the estimation of $z \sim 2$ black hole masses using galaxy-black hole correlations. In order to accurately characterize the uncertainties of our estimated black hole masses, we calculate black hole masses from host galaxy stellar masses via Monte Carlo simulations that account for the uncertainties in each quantity involved in the calculation {\changed --- stellar mass, redshift and a possible evolutionary correction of up to a factor of 3 ---} as well as the scatter in the black hole mass-host stellar mass correlation of \citet{haringrix04}. We simulate $10^5$ observations via Gaussian-random sampling within the uncertainties for each source and report the median black hole masses and $1 \sigma$ widths of the distributions of each source as our uncertainties.

Table \ref{datatable} gives the black hole mass estimates for each of our sources. We find that the black hole masses span a large range, from $\MBHminr$ to $\MBHmaxr$, with median and mean masses of $\MBHmedr$ and $\MBHavgr$, respectively. Typical uncertainties are $-0.7$ and $+1.0$ dex, with asymmetries in the uncertainties primarily due to uncertainties in the host stellar masses, which are also calculated by Monte Carlo methods \citep{kriek09a}. {\changed As a sanity check, we determined the black hole mass of X577 using the FWHM of the \mgii\ broad line and the continuum luminosity at 3000 \AA\ \citep{mclure04b}. The latter is uncertain in the spectrum \citep{szokoly04}, but the calculated mass of $1.0^{+0.8}_{-0.3} \times 10^8~\msol$ is fully consistent with that determined using the Monte Carlo method described above.}

In addition to black hole masses, bolometric (total) luminosities are useful quantities because they allow for an estimate of the mass accretion rate onto the black hole (modulo the radiative efficiency). \citet{simmons11} have shown that the model-dependent bolometric corrections to the hard X-ray luminosity of \citet{treister09} are in very good agreement with bolometric luminosities recovered for obscured AGN by leveraging the observed point-source luminosities to extract the SEDs from the far-infrared to X-ray of each AGN. We therefore use the absorption-corrected X-ray luminosity of each source, along with the bolometric corrections of \citet{treister09}, to estimate the bolometric luminosities of each source. Given the X-ray luminosities in the sample, the bolometric correction ranges from $25$ to $76$.

\begin{figure}
\figurenum{5}
\epsscale{1.15}
\plotone{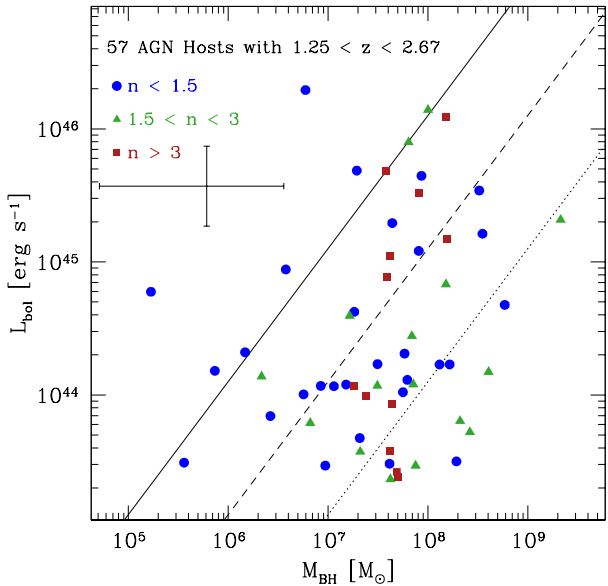}\label{Mbh_lbol}
\caption{Bolometric luminosity vs. black hole mass for AGN in disk-dominated (blue circles), intermediate (green triangles) and strong-bulge (red squares) host galaxies at $1.25 < z < 2.7$ in this sample. {\changed Median error bars are shown at left beneath the legend.} Lines shown indicate constant Eddington ratio: $L = L_{Edd}$ (solid line), $L = 0.1 L_{Edd}$ (dashed line), and $L = 0.01 L_{Edd}$ (dotted line). Growing black holes at these redshifts span three orders of magnitude of growth rate, from Eddington-limited (and possibly super-Eddington) rates to highly sub-Eddington. One-third of the sample has $L/L_{Edd} \leq 0.01$; another $\sim 30$\% have $0.01 \leq L/L_{Edd} \leq 0.1$. Bolometric luminosity is uncorrelated with host galaxy morphology.}
\end{figure}

With estimates of both bolometric luminosity and black hole mass, we explore the growth rates of the black holes in the sample in Figure \ref{Mbh_lbol}. The growth rates of moderate-luminosity AGN at $z \sim 2$ span several orders of magnitude, from super-Eddington accretion to $10^{-3} L_{Edd}$, with average and median $L/L_{Edd}$ values of $0.06$ and $0.05$, respectively. One-quarter of the sample (19 of 57 AGN) has $L/L_{Edd} < 0.01$; 21 AGN have $0.01 \leq L/L_{Edd} \leq 0.1$. The sample cannot therefore be characterized by a single growth rate/Eddington ratio, but {\changed at least half of the sample is growing at less than 10\% of the Eddington (maximum) rate (96\% confidence level).}

\section{Discussion}\label{discussion}

Because moderate-luminosity AGN far outnumber more luminous objects like quasars, they make up a large fraction of black hole growth in the universe \citep{hasinger05}. They therefore represent the most common mode of black hole feeding that we can observe, and their feedback mechanism(s) represents the typical means of black hole-galaxy co-evolution.

The host galaxies of the sample are generally characterized by strong disk contributions. Fully half of the sample of host galaxies is consistent with having less than 20\% bulge contribution, and a further one-quarter of the AGN are hosted in galaxies with between 20 and 65\% of their light from a bulge. Very few host galaxies in the sample are unambiguously consistent with the label ``bulge-dominated'': \citet{simmons08} showed that even a host galaxy with a fitted S\'ersic index of $n\geq4$ (traditionally associated with a pure bulge or elliptical; there are 8 in the sample) may have a strong disk contribution (up to 45\% in some cases). Non-detection of a bulge, on the other hand, is a reliable result according to multiple sets of host galaxy simulations  \citep{sanchez04,simmons08, gabor09,pierce10a}.

That 75\% of moderate-luminosity $z \sim 2$ AGN hosts in the CDF-S have strong disk components agrees quantitatively with the pilot study of \citet{schawinski11}. Using visual classifications, \citet{kocevski11} find a slightly lower fraction of disks (51\%). However, visual classifications of host galaxies with even faint central nuclear emission may be biased toward recovering more bulge-like host morphologies \citep{simmons12a}. Given that 90\% of moderate-luminosity AGN+hosts at $z \sim 2$ have detected central point sources, the disk fraction of the \citeauthor{kocevski11} sample should be considered a lower limit.

The high fraction of disky morphologies in moderate-luminosity AGN hosts at $z \sim 2$ (Figure \ref{nhist}) means the majority of the host galaxies in the sample have assembled without building up prominent bulges. This implies that major mergers, which tend to destroy disks and lead to elliptical or bulge-dominated morphologies \citep[e.g.,][]{toomre77,walker96,martig12}, play a small role in the growth of these galaxies and their central black holes. 

{\changed Note that, while a disk \emph{may} be able to re-form after a gas-rich merger \citep[e.g.,][]{hopkins09c}, this merger process still strongly favors the formation of a bulge. Such a formation history is unlikely to be typical for the half of the sample with $n \leq 1.5$, and even less likely for the one-third of the sample with $n \leq 1$, which is consistent at these redshifts with a bulge fraction of less than $10$\% \citep{simmons08}. Additionally, the majority of the sample does not show the signatures of a significant recent merger \citep[the brighter features of such signatures could be detected in images at this depth;][]{schawinski12}. Thus gas-rich mergers that enable disks to re-form yet fail to form significant bulges either occurred at $z \gg 2$, so that by $z \sim 2$ they have relaxed to the point where they are not detected in deep rest-frame optical \emph{HST} images, or they do not occur in moderate-luminosity AGN.}

Those galaxies that do appear to be highly disturbed also do not host the highest-luminosity AGN, as might be expected if the processes that disrupted the host also fed large amounts of material onto the black hole. Black hole-galaxy co-evolution scenarios in which major mergers trigger accretion onto the central supermassive black hole \citep{sanders88, dimatteo05, hopkins06}, leading from an Eddington-limited quasar phase in a merger remnant host to a longer phase of slow black hole growth with a moderate- or low-luminosity AGN hosted by an elliptical galaxy, are inconsistent with the morphologies of moderate-luminosity AGN hosts at this early epoch.

Disk-dominated hosts are more likely to be observed as blue than bulges: although there are a substantial number of intrinsically red disks in the sample, uncorrected colors of the sub-samples with $n < 1.5$ and $n > 3$ are different at the 99\% level, according to a K-S test. However, this difference disappears once the dust correction is applied, primarily because several of the strong-bulge hosts appear to be intrinsically blue, but dust-reddened. Blue early-type galaxies have also been seen for AGN hosts in other surveys at lower redshifts  \citep{sanchez04,pierce10b,cardamone10}.

However, the rest-frame $U-V$ colors of $z \sim 2$ AGN hosts are not uniformly consistent with the ``blue cloud'' population at the same redshift, as expected from galaxy formation simulations where mergers trigger black hole growth and a burst of star formation \citep{hopkins06b,somerville08}. Even after a dust correction is applied, $\sim 50$\% of host galaxies are red, in both redshift bins.


Merger scenarios tying the evolutionary stage of a post-merger to the decaying accretion rate of its central supermassive black hole also predict a correlation between AGN luminosity and host galaxy morphology. However, $z\sim 2$ AGN bolometric luminosity does not depend on host galaxy morphology: as seen in Figure \ref{Mbh_lbol}, black holes growing at or near the Eddington limit are found in disk-dominated, intermediate, and strong-bulge host galaxies, as are the slowest-growing black holes in the sample. 

\begin{figure}
\figurenum{6}
\epsscale{1.15}
\plotone{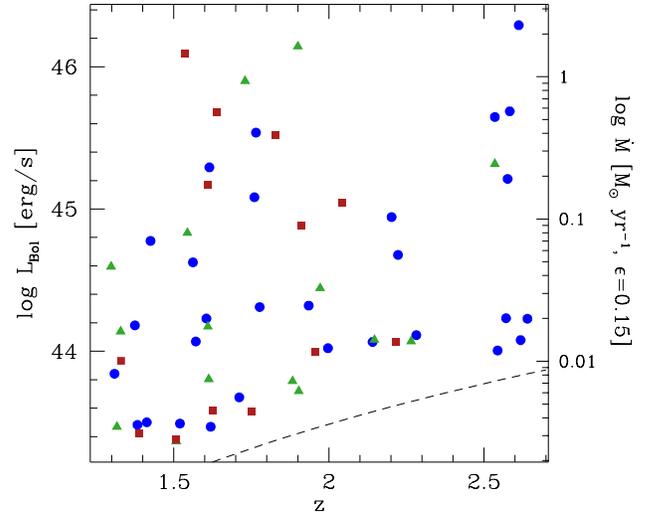}\label{Lbol_Mdot_z}
\caption{
Bolometric luminosity vs. redshift for moderate-luminosity AGN in the GOODS-S/CANDELS field. Given the conversion of the X-ray flux limit of the \citet{xue11} survey (dashed line) to a bolometric limit, 
our sources fill the observable area. The bolometric luminosity is converted into a mass accretion rate on the right vertical axis, assuming a radiative efficiency of $\epsilon=0.15$ \citep{elvis02}.
}
\end{figure}

Nor does the bolometric luminosity depend on redshift: black holes at all redshifts in the sample grow at all rates we are able to detect. In Figure \ref{Lbol_Mdot_z} we show the bolometric luminosity versus redshift for the moderate-luminosity AGN in this sample. We also plot the flux limit of the sample, which defines a redshift-dependent lower limit to the bolometric luminosity. 

Figure \ref{Lbol_Mdot_z} shows that the observed AGN populate every part of the parameter space that we do not select against. The range of AGN luminosities sampled implies a range of mass accretion rates spanning over two orders of magnitude. For an assumed radiative efficiency of $\epsilon = 0.15$ \citep{elvis02}, the black holes in the sample are growing at a rate of no more than $\sim 2 M_{\odot}$ per year for the highest-luminosity black hole. Lower accretion rates are more typical of the sample, with the average and median values being $0.03$ and $0.02 M_{\odot}$ yr$^{-1}$, respectively, and the lowest-luminosity AGN in the sample accreting at $0.003 M_{\odot}$ yr$^{-1}$. 

Such characteristically low mass accretion rates imply that slow- and moderate-growth periods of a typical supermassive black hole do not require large, bulk flows of matter into the central region of the host galaxy, as one might expect from merger-triggered gas infall. Indeed, the median rate of $0.02 M_{\odot}$ yr$^{-1}$ is roughly consistent with expected stellar mass loss from a passively evolving stellar population (\citeauthor{ciotti91} \citeyear{ciotti91}; for a recent treatment, see \citeauthor{ciotti12} \citeyear{ciotti12}).

\section{Conclusions}

We describe 57 AGN host galaxies from $1.25 \leq z \leq 2.67$ in the GOODS-South field selected from the 4~Ms X-ray catalogs of \citet{xue11}. Using  \emph{HST} WFC3/IR $F125W$ and $F160W$ images from CANDELS, we perform parametric morphological fits in the rest-frame $B$ band. We additionally calculate stellar masses, black hole masses, bolometric AGN luminosities, and rest-frame host galaxy $(U-V)$ colors for each source.

Our major results are as follows:

\begin{itemize}
\item Host galaxy morphologies span a range from disk-dominated to bulge-dominated, but approximately 50\% of the sample is unambiguously disk-dominated, and another $\sim25\%$ of the host sample have fitted S\'ersic indices consistent with strong (if not dominant) disk components.

\item Host morphologies are statistically indistinguishable between two redshift bins of approximately equal time intervals ($\Delta t \approx 1.3 \mathrm{~Gyr;~} 1.25 < z < 1.76$ and $1.76 < z < 2.67$).

\item Uncorrected host galaxy colors between the two redshift bins do appear different, with blue colors more likely at higher redshifts, but the significance of this difference is reduced (from $\approx 97\%$ to $\approx 74$\% according to a K-S test) once a dust correction is applied.

\item We detect central point sources in the rest-frame $B$ band to at least $3 \sigma$ in 90\% of sources. The {\changed nuclei are typically faint compared to the host galaxy; the} majority of detections are greater than $5 \sigma$.

\item Central black hole masses, estimated from stellar masses, span a wide range (from $10^5-10^9 M_{\odot}$), with a median mass of $\MBHmedr$.

\item AGN bolometric luminosities range from $2 \times 10^{43}$ to $2 \times 10^{46}$~erg~s$^{-1}$; this is limited only by the flux limits of the 4~Ms \emph{Chandra} data and expected detection numbers based on the AGN luminosity function at these redshifts \citep{croom04}. Essentially, we find AGN accreting at all levels we are capable of detecting, with no redshift trend.

\item AGN bolometric luminosity is uncorrelated with host stellar mass, color or morphology, in contrast to the predictions of some models of merger-driven accretion.

\end{itemize} 

Moderate-luminosity AGN, which collectively comprise a substantial fraction of total black hole growth at these redshifts, span several orders of magnitude in growth rate and are hosted by galaxies with a range of star formation rates, as indicated by a range of dust-corrected colors. While their host galaxies also span a range of morphologies, strong disks dominate. That fully half of growing black holes at these luminosities are hosted in galaxies with a (conservative) minimum of 80\% of their optical light coming from an undisturbed disk likely indicates the importance of secular processes and minor mergers (that do not produce a significant bulge) in the co-evolution of black holes and galaxies.

\acknowledgments

The authors wish to thank to C. Peng for making \galfit\ publicly available, and for many enlightening discussions.  
The JavaScript Cosmology Calculator \citep{wright06} and TOPCAT \citep{taylor05} were used while preparing this paper. 
We thank M. Vestergaard for kindly providing templates for Fe pseudo-continuum emission \citep{vestergaard01}, which were used to fit the spectrum of X577.
BDS acknowledges support from Worcester College, Oxford, and from NASA through grant HST-AR-12638.01-A from the Space Telescope Science Institute, which is operated by the Association of Universities for Research in Astronomy under NASA contract NAS 5-26555.  
CMU acknowledges support from Chandra Grant SP1-12004X and Hubble Archival Grant SP1-12004X.
KS gratefully acknowledges support from Swiss National Science Foundation Grant PP00P2\_138979/1 and from NASA through Einstein Postdoctoral Fellowship grant number PF9-00069, issued by the Chandra X-ray Observatory Center, which is operated by the Smithsonian Astrophysical Observatory for and on behalf of NASA under contract NAS8-03060.
This work is based on observations taken by the CANDELS Multi-Cycle Treasury Program with the NASA/ESA HST, which is operated by the Association of Universities for Research in Astronomy, Inc., under NASA contract NAS5-26555.



\LongTables

\begin{deluxetable*}{lccccccccccr}
\tablecaption{{\changed GOODS AGN and Host Galaxy Properties At $1.25 < z < 2.67$.}}
\tablehead{
\multicolumn{1}{c}{}  &  \multicolumn{2}{c}{ Optical Position} & \multicolumn{1}{c}{}  & \multicolumn{3}{c}{Rest-frame $B$} & \multicolumn{2}{c}{$\log L$ [erg/s]} & \multicolumn{2}{c}{$\log M$ [M$_\odot$]} &\\ \\
\colhead{ID\tablenotemark{a}} &\colhead{RA} &\colhead{DEC} &\colhead{$z$}  &\colhead{S\'ersic $n$} &\colhead{$\log L_{\rm PS}/L_{\rm Host}$} & \colhead{PS Signif. ($\sigma$)} &\colhead{$L_X$\tablenotemark{b}} &\colhead{$L_{bol}$} &\colhead{Host $M_*$} & \colhead{$M_{BH}$} & \colhead{{\changed Notes}\tablenotemark{c}}
}
\tablecolumns{11}
\startdata
  121 &     53.026794 &    -27.765278 &    1.33 & $ 2.13^{+ 0.34}_{- 0.34}$ & $ -1.15 \pm 0.77$ & 5 &    42.61 &    44.14 & $     9.12_{-  0.51}^{+  1.98}$ & $     6.34_{-  0.70}^{+  2.13}$ & \\ 
  137 &     53.033417 &    -27.782642 &    2.61 & $ 0.88^{+ 0.34}_{- 0.34}$ & $ -0.75 \pm 0.73$ & $>5$ &    44.41 &    46.29 & $     9.49_{-  0.66}^{+  1.35}$ & $     6.78_{-  0.84}^{+  1.47}$ & \\ 
  155 &     53.040958 &    -27.836111 &    1.93 & $ 0.34^{+ 0.27}_{- 0.27}$ & $ -0.05 \pm 0.24$ & $>5$ &    42.76 &    44.32 & $     8.97_{-  0.56}^{+  1.30}$ & $     6.17_{-  0.74}^{+  1.41}$ & A, M \\ 
  163 &     53.044960 &    -27.774416 &    1.61 & $ 1.56^{+ 0.29}_{- 0.29}$ & $ -1.41 \pm 0.62$ & $>5$ &    42.64 &    44.17 & $    11.20_{-  0.26}^{+  0.47}$ & $     8.61_{-  0.37}^{+  0.53}$ & C\\ 
  184 &     53.052288 &    -27.774778 &    1.61 & $ 1.40^{+ 0.33}_{- 0.33}$ & $ -1.18 \pm 0.77$ & $>5$ &    42.69 &    44.23 & $    10.84_{-  0.18}^{+  0.43}$ & $     8.22_{-  0.31}^{+  0.47}$ & \\ 
  185 &     53.052372 &    -27.827280 &    2.15 & $ 1.54^{+ 0.44}_{- 0.44}$ & $ -0.84 \pm 0.73$ & $>5$ &    42.56 &    44.08 & $    10.52_{-  1.35}^{+  1.53}$ & $     7.85_{-  1.54}^{+  1.72}$ & A \\ 
  199 &     53.057919 &    -27.833555 &    2.54 & $ 1.79^{+ 0.26}_{- 0.26}$ & $ -1.61 \pm 0.48$ & $>5$ &    43.60 &    45.32 & $    11.81_{-  0.63}^{+  0.93}$ & $     9.33_{-  0.76}^{+  1.04}$ & \\ 
  202 &     53.058956 &    -27.819613 &    2.14 & $ 1.01^{+ 0.36}_{- 0.36}$ & $ -0.64 \pm 0.71$ & $>5$ &    42.55 &    44.07 & $     9.83_{-  0.82}^{+  0.81}$ & $     7.06_{-  0.95}^{+  0.92}$ & C\\ 
  205 &     53.060123 &    -27.853054 &    1.54 & $ 2.83^{+ 0.30}_{- 0.30}$ & $ -0.95 \pm 0.74$ & $>5$ &    43.19 &    44.83 & $    10.87_{-  0.51}^{+  0.41}$ & $     8.18_{-  0.59}^{+  0.50}$ & A\\ 
  211 &     53.061958 &    -27.851082 &    1.83 & $ 6.76^{+ 0.84}_{- 0.84}$ & $ -1.29 \pm 0.71$ & $>5$ &    43.76 &    45.52 & $    10.53_{-  0.39}^{+  1.12}$ & $     7.91_{-  0.54}^{+  1.20}$ & \\ 
  221 &     53.065670 &    -27.879000 &    1.97 & $ 2.20^{+ 0.29}_{- 0.29}$ & $ -0.56 \pm 0.70$ & 5 &    42.87 &    44.44 & $    10.45_{-  0.30}^{+  1.01}$ & $     7.84_{-  0.46}^{+  1.08}$ & \\ 
  225 &     53.066292 &    -27.800610 &    1.52 & $ 0.90^{+ 1.39}_{- 0.86}$ & \nodata & $<1$ &    42.07 &    43.49 & $     8.51_{-  1.61}^{+  1.73}$ & $     5.56_{-  1.83}^{+  1.94}$ & \\ 
  226 &     53.066837 &    -27.816639 &    1.41 & $ 0.58^{+ 0.32}_{- 0.32}$ & $ -1.13 \pm 0.76$ & $>5$ &    42.08 &    43.50 & $    11.00_{-  1.14}^{+  0.60}$ & $     8.29_{-  1.26}^{+  0.73}$ & C\\ 
  231 &     53.068459 &    -27.866526 &    2.20 & $ 1.18^{+ 0.64}_{- 0.64}$ & $ 0.05 \pm 0.20$ & $>5$ &    43.28 &    44.94 & $     9.35_{-  0.81}^{+  1.33}$ & $     6.58_{-  0.98}^{+  1.46}$ & \\ 
  242 &     53.071621 &    -27.769861 &    1.33 & $ 8.74^{+ 0.47}_{- 0.47}$ & \nodata & 2 &    42.44 &    43.93 & $    10.39_{-  0.80}^{+  0.80}$ & $     7.64_{-  0.91}^{+  0.91}$ & A\\ 
  244 &     53.072121 &    -27.819002 &    2.28 & $ 0.76^{+ 0.20}_{- 0.20}$ & \nodata & 1 &    42.59 &    44.11 & $    10.36_{-  0.39}^{+  1.69}$ & $     7.79_{-  0.59}^{+  1.80}$ & \\ 
  247 &     53.074337 &    -27.869667 &    1.88 & $ 2.22^{+ 0.30}_{- 0.30}$ & $ -1.49 \pm 0.57$ & $>5$ &    42.32 &    43.79 & $     9.58_{-  1.04}^{+  1.89}$ & $     6.82_{-  1.24}^{+  2.08}$ & C\\ 
  257 &     53.076458 &    -27.848778 &    1.54 & $ 3.31^{+ 0.33}_{- 0.33}$ & $ -1.08 \pm 0.76$ & $>5$ &    44.24 &    46.09 & $    10.82_{-  0.17}^{+  0.33}$ & $     8.18_{-  0.29}^{+  0.38}$ & A\\ 
  288 &     53.086922 &    -27.873026 &    1.42 & $ 0.79^{+ 1.54}_{- 0.75}$ & $ -0.90 \pm 0.85$ & $>5$ &    43.14 &    44.78 & $     8.21_{-  0.81}^{+  0.97}$ & $     5.23_{-  0.96}^{+  1.09}$ & \\ 
  293 &     53.088795 &    -27.850554 &    1.57 & $ 0.53^{+ 0.26}_{- 0.26}$ & $ -0.64 \pm 0.71$ & $>5$ &    42.55 &    44.07 & $     9.66_{-  0.33}^{+  1.03}$ & $     6.93_{-  0.49}^{+  1.10}$ & C\\ 
  296 &     53.090752 &    -27.782528 &    1.56 & $ 0.39^{+ 0.30}_{- 0.30}$ & $ -0.89 \pm 0.74$ & $>5$ &    43.02 &    44.62 & $     9.99_{-  0.63}^{+  1.37}$ & $     7.26_{-  0.78}^{+  1.49}$ & \\ 
  301 &     53.092419 &    -27.803249 &    1.76 & $ 1.27^{+ 0.32}_{- 0.32}$ & $ -1.34 \pm 0.67$ & $>5$ &    43.40 &    45.08 & $    10.63_{-  1.26}^{+  0.89}$ & $     7.91_{-  1.40}^{+  1.04}$ & \\ 
  305 &     53.093834 &    -27.801357 &    1.91 & $ 3.63^{+ 1.34}_{- 1.34}$ & $ -0.98 \pm 0.80$ & 3 &    43.24 &    44.89 & $    10.41_{-  2.75}^{+  0.63}$ & $     7.59_{-  3.00}^{+  0.82}$ & \\ 
  308 &     53.094002 &    -27.767834 &    1.73 & $ 2.65^{+ 0.85}_{- 0.85}$ & $ -0.60 \pm 0.71$ & $>5$ &    44.08 &    45.90 & $    10.57_{-  1.04}^{+  0.50}$ & $     7.81_{-  1.13}^{+  0.63}$ & C\\ 
  310 &     53.094086 &    -27.804220 &    2.54 & $ 0.99^{+ 0.29}_{- 0.29}$ & $ -0.89 \pm 0.74$ & $>5$ &    43.87 &    45.65 & $    10.52_{-  0.68}^{+  1.60}$ & $     7.94_{-  0.86}^{+  1.75}$ & \\ 
  325 &     53.099957 &    -27.808554 &    2.22 & $ 8.78^{+ 1.19}_{- 1.19}$ & $ -0.62 \pm 0.71$ & $>5$ &    42.55 &    44.06 & $     9.92_{-  0.39}^{+  1.10}$ & $     7.26_{-  0.56}^{+  1.18}$ & \\ 
  337 &     53.103535 &    -27.847334 &    2.27 & $ 1.97^{+ 0.33}_{- 0.33}$ & $ -1.25 \pm 0.74$ & $>5$ &    42.55 &    44.07 & $    10.13_{-  0.37}^{+  1.07}$ & $     7.49_{-  0.54}^{+  1.15}$ & C\\ 
  360 &     53.108372 &    -27.797722 &    1.71 & $ 0.11^{+ 0.28}_{- 0.07}$ & $ -0.78 \pm 0.72$ & $>5$ &    42.22 &    43.68 & $    10.04_{-  0.65}^{+  1.10}$ & $     7.32_{-  0.79}^{+  1.21}$ & A, M\\ 
  389 &     53.119244 &    -27.765888 &    1.63 & $ 3.35^{+ 0.54}_{- 0.54}$ & \nodata & $<1$ &    42.14 &    43.58 & $    10.32_{-  0.72}^{+  1.20}$ & $     7.62_{-  0.87}^{+  1.32}$ & \\ 
  394 &     53.120205 &    -27.798887 &    1.38 & $ 0.73^{+ 0.31}_{- 0.31}$ & $ -1.08 \pm 0.76$ & $>5$ &    42.06 &    43.48 & $    10.31_{-  0.25}^{+  0.62}$ & $     7.62_{-  0.38}^{+  0.67}$ & A, M\\ 
  428 &     53.129620 &    -27.827780 &    1.51 & $ 1.81^{+ 0.59}_{- 0.59}$ & $ -0.70 \pm 0.74$ & $>5$ &    41.96 &    43.37 & $    10.35_{-  0.57}^{+  0.72}$ & $     7.62_{-  0.68}^{+  0.81}$ & \\ 
  435 &     53.131165 &    -27.773222 &    2.22 & $ 1.27^{+ 0.24}_{- 0.24}$ & $ -0.44 \pm 0.63$ & $>5$ &    43.06 &    44.68 & $    11.43_{-  0.92}^{+  0.35}$ & $     8.77_{-  0.99}^{+  0.51}$ & A\\ 
  436 &     53.131287 &    -27.841389 &    1.61 & $ 2.77^{+ 0.35}_{- 0.35}$ & $ -1.84 \pm 0.34$ & $>5$ &    42.33 &    43.80 & $    10.94_{-  0.37}^{+  0.70}$ & $     8.32_{-  0.49}^{+  0.77}$ & C\\ 
  437 &     53.131454 &    -27.814999 &    1.78 & $ 0.21^{+ 0.18}_{- 0.17}$ & \nodata & $<1$ &    42.75 &    44.31 & $    10.46_{-  0.82}^{+  1.03}$ & $     7.77_{-  0.96}^{+  1.16}$ & C\\ 
  451 &     53.137287 &    -27.844805 &    2.62 & $ 0.95^{+ 0.31}_{- 0.31}$ & $ -1.05 \pm 0.76$ & 4 &    42.56 &    44.08 & $     9.84_{-  0.26}^{+  0.66}$ & $     7.18_{-  0.44}^{+  0.71}$ & A\\ 
  462 &     53.140289 &    -27.797556 &    1.39 & $ 7.47^{+ 1.00}_{- 1.00}$ & $ -0.66 \pm 0.71$ & $>5$ &    42.01 &    43.42 & $    10.44_{-  0.58}^{+  0.48}$ & $     7.69_{-  0.67}^{+  0.57}$ & \\ 
  463 &     53.140999 &    -27.766834 &    1.90 & $ 2.47^{+ 0.27}_{- 0.27}$ & $ -0.63 \pm 0.70$ & $>5$ &    42.26 &    43.72 & $    11.00_{-  0.30}^{+  0.67}$ & $     8.42_{-  0.43}^{+  0.73}$ & C\\ 
  464 &     53.141037 &    -27.755833 &    2.04 & $ 6.37^{+ 0.81}_{- 0.81}$ & $ -1.27 \pm 0.73$ & $>5$ &    43.37 &    45.04 & $    10.36_{-  1.34}^{+  0.92}$ & $     7.62_{-  1.49}^{+  1.07}$ & A, M\\ 
  482 &     53.146084 &    -27.780027 &    2.64 & $ 1.20^{+ 0.34}_{- 0.34}$ & $ -1.19 \pm 0.78$ & 5 &    42.69 &    44.23 & $    10.70_{-  0.85}^{+  1.60}$ & $     8.12_{-  1.03}^{+  1.74}$ & \\ 
  490 &     53.148834 &    -27.821222 &    2.58 & $ 0.51^{+ 0.29}_{- 0.29}$ & $ -0.92 \pm 0.74$ & $>5$ &    43.51 &    45.21 & $    11.11_{-  0.89}^{+  1.26}$ & $     8.55_{-  1.05}^{+  1.41}$ & A\\ 
  493 &     53.149918 &    -27.814083 &    1.31 & $ 0.72^{+ 0.28}_{- 0.28}$ & $ -0.84 \pm 0.73$ & $>5$ &    42.36 &    43.84 & $     9.29_{-  0.10}^{+  0.14}$ & $     6.42_{-  0.23}^{+  0.24}$ & A, M\\ 
  498 &     53.150707 &    -27.843723 &    1.61 & $ 3.54^{+ 0.33}_{- 0.33}$ & $ -1.02 \pm 0.75$ & $>5$ &    43.47 &    45.17 & $    10.82_{-  0.37}^{+  0.72}$ & $     8.19_{-  0.49}^{+  0.79}$ & A, M\\ 
  501 &     53.150791 &    -27.774445 &    2.57 & $ 0.74^{+ 0.32}_{- 0.32}$ & $ -1.13 \pm 0.76$ & $>5$ &    42.69 &    44.23 & $    10.14_{-  0.65}^{+  1.41}$ & $     7.50_{-  0.82}^{+  1.55}$ & C\\ 
  504 &     53.151455 &    -27.825945 &    1.51 & $ 6.82^{+ 0.70}_{- 0.70}$ & $ -1.07 \pm 0.76$ & $>5$ &    41.98 &    43.38 & $    10.39_{-  0.48}^{+  0.85}$ & $     7.70_{-  0.60}^{+  0.93}$ & \\ 
  517 &     53.157288 &    -27.833723 &    1.62 & $ 0.31^{+ 0.29}_{- 0.27}$ & $ -0.87 \pm 0.73$ & $>5$ &    42.05 &    43.47 & $     9.73_{-  0.48}^{+  0.92}$ & $     6.97_{-  0.62}^{+  1.01}$ & A, M\\ 
  525 &     53.160500 &    -27.776361 &    2.54 & $ 0.80^{+ 0.27}_{- 0.27}$ & $ -0.66 \pm 0.71$ & $>5$ &    42.50 &    44.00 & $     9.47_{-  0.59}^{+  1.31}$ & $     6.76_{-  0.77}^{+  1.42}$ & A\\ 
  545 &     53.165207 &    -27.785999 &    1.32 & $ 2.30^{+ 0.38}_{- 0.38}$ & $ -1.21 \pm 0.77$ & $>5$ &    42.05 &    43.47 & $    10.54_{-  0.21}^{+  0.60}$ & $     7.88_{-  0.34}^{+  0.64}$ & \\ 
  549 &     53.165585 &    -27.769861 &    1.76 & $ 1.03^{+ 0.33}_{- 0.33}$ & $ -1.20 \pm 0.77$ & $>5$ &    43.78 &    45.54 & $    11.16_{-  0.70}^{+  0.54}$ & $     8.51_{-  0.79}^{+  0.65}$ & C\\ 
  552 &     53.166874 &    -27.798834 &    2.00 & $ 0.83^{+ 0.32}_{- 0.32}$ & $ -1.14 \pm 0.77$ & $>5$ &    42.51 &    44.02 & $    10.38_{-  0.31}^{+  0.87}$ & $     7.75_{-  0.46}^{+  0.93}$ & C\\ 
  557 &     53.170666 &    -27.741056 &    1.30 & $ 2.07^{+ 0.28}_{- 0.28}$ & $ -0.77 \pm 0.72$ & 5 &    42.99 &    44.60 & $     9.98_{-  0.47}^{+  0.75}$ & $     7.22_{-  0.59}^{+  0.83}$ & \\ 
  575 &     53.179371 &    -27.812611 &    1.64 & $ 5.53^{+ 0.51}_{- 0.51}$ & $ -0.94 \pm 0.74$ & $>5$ &    43.90 &    45.68 & $    10.27_{-  0.31}^{+  0.66}$ & $     7.58_{-  0.44}^{+  0.72}$ & A\\ 
  577 &     53.180168 &    -27.820696 &    1.90 & $ 2.79^{+ 0.29}_{- 0.29}$ & $ -0.71 \pm 0.71$ & $>5$ &    44.29 &    46.14 & $    10.62_{-  0.23}^{+  0.57}$ & $     8.00_{-  0.37}^{+  0.62}$ & \\ 
  579 &     53.181038 &    -27.817280 &    1.75 & $ 5.50^{+ 0.31}_{- 0.31}$ & \nodata & $<1$ &    42.14 &    43.57 & $    10.03_{-  0.53}^{+  1.12}$ & $     7.32_{-  0.68}^{+  1.22}$ & \\ 
  589 &     53.185040 &    -27.819805 &    1.95 & $ 3.86^{+ 0.54}_{- 0.54}$ & $ -1.36 \pm 0.66$ & $>5$ &    42.49 &    43.99 & $    10.10_{-  1.32}^{+  1.51}$ & $     7.38_{-  1.51}^{+  1.69}$ & A, M\\ 
  593 &     53.185875 &    -27.810055 &    2.58 & $ 1.37^{+ 0.77}_{- 0.77}$ & $ -0.18 \pm 0.38$ & 4 &    43.90 &    45.69 & $    10.06_{-  1.28}^{+  0.80}$ & $     7.29_{-  1.42}^{+  0.96}$ & \\ 
  625 &     53.198872 &    -27.844002 &    1.62 & $ 1.10^{+ 0.31}_{- 0.31}$ & $ -1.06 \pm 0.76$ & $>5$ &    43.58 &    45.29 & $    10.30_{-  0.44}^{+  1.37}$ & $     7.64_{-  0.59}^{+  1.49}$ & C\\ 
  656 &     53.218037 &    -27.761694 &    1.38 & $ 1.26^{+ 0.28}_{- 0.28}$ & $ -0.83 \pm 0.73$ & $>5$ &    42.65 &    44.18 & $     8.80_{-  0.23}^{+  0.23}$ & $     5.87_{-  0.34}^{+  0.33}$ & A
\enddata
\tablenotetext{a}{X-ray IDs from \citet{xue11}}

\tablenotetext{b}{X-ray luminosity from $0.5 - 8$ keV}

\tablenotetext{c}{C: Clumpy; A: Asymmetric; M: Merger. For details, see Section \ref{sec:morph}.}
\label{datatable}
\end{deluxetable*}

\bibliographystyle{apj}

\end{document}